\documentclass[prd,reprint,showpacs,showkeys]{revtex4}
\usepackage{amsfonts}
\usepackage{mdframed}
\usepackage{amssymb}
\usepackage{amsmath}
\usepackage{graphicx}
\usepackage[font={footnotesize,it}]{caption}

\begin{document}

\title{The Stability of Asymmetric Cylindrical Thin-Shell Wormholes}
\author{S. Danial Forghani}
\email{danial.forghani@emu.edu.tr}
\author{S. Habib Mazharimousavi}
\email{habib.mazhari@emu.edu.tr}
\author{M. Halilsoy}
\email{mustafa.halilsoy@emu.edu.tr}
\affiliation{Department of Physics, Faculty of Arts and Sciences, Eastern Mediterranean
University, Famagusta, North Cyprus via Mersin 10, Turkey}

\begin{abstract}
In continuation of a preceding work on introducing asymmetric thin-shell
wormholes as an emerging class of traversable wormholes within the context,
this time cylindrically symmetric spacetimes are exploited to construct such
wormholes. Having established a generic formulation, first the Linet-Tian
metric generally, and then the cosmic string metric and a black string
metric in greater details are studied as constructing blocks of cylindrical
asymmetric thin-shell wormholes. The corresponding wormholes are
investigated within the linearized stability analysis framework to firstly,
demonstrate that they can exist from the mechanical stability point of view,
and secondly, indicate the correlation between the stability and symmetry in
each case, if there is any at all. From here, we have extracted a pattern
for the way stability changes with the asymmetry degree for the two
examples; however, it was observed that the symmetric state is not the most
neither the less stable state. There are also some side results: It was
learned that any cylindrical thin-shell wormhole made of two cosmic string
universes cannot be supported by a barotropic equation of state.
Furthermore, as another side outcome, it was perceived that the radius
dependency of the so-called variable equation of state, which is used all
over this article, has a great impact on the mechanical stability of the
cylindrical asymmetric thin-shell wormholes studied in this brief.
\end{abstract}

\maketitle

\section{Introduction}

Cylindrically symmetric wormholes \cite{CW} and cylindrically symmetric
thin-shell wormholes (TSWs) \cite{CTSW} have been considered in literature.
Also in a previous study, we introduced the concept of asymmetric thin-shell
wormholes (ATSWs) with explicit constructions as particular examples in
spherical coordinate system \cite{Forghani}. Briefly, any thin-shell
wormhole (TSW), constructed in an inhomogeneous or non-isotropic bulk
spacetime will fail to satisfy the mirror symmetry in the embedding diagram
and therefore it can be dubbed as an ATSW. Apart from the concept of ATSW,
asymmetric wormholes have also been considered in the literature \cite%
{AW1,AW2,AW3}. We expect changes under such a reduced symmetry as
encountered in different branches of theoretical physics. Are there emergent
physical quantities as a result of such a broken mirror symmetry?

The spherically symmetric spacetimes are one degree higher symmetric systems
compared to the cylindrically symmetric ones. This can easily be visualized
by comparing a sphere and a cylinder. In analogy, an axially symmetric space
can also be categorized in a less symmetric configuration in comparison with
the spherically symmetric ones. Rotation of a spherically symmetric
spacetime is known to lose its spherical symmetry and it transforms into a
stationary symmetric one. This is exactly how rotation transforms the
Schwarzschild spacetime into the stationary Kerr spacetime.

The main disadvantage of a cylindrical system is the occurrence of a
non-compact direction, i.e. the $z-$axis, so that we cannot mention of an
asymptotic flatness in such a spacetime. Only for a slice, say $z=$const.,
we can carry the radial coordinate to spatial infinity and discuss of
asymptotic flatness in a restricted sense. The singularity structure of a
cylindrically symmetric system is also different from a compact spherical
system.

Another aspect which makes the subject matter of the present article is the
stability analysis of an ATSW in a cylindrically symmetric spacetime. In
other words, we shall construct such explicit wormholes first and consider
their stability under radial perturbations. Unlike the stability analysis of
a localized spherical system, in a cylindrical system perturbation must be
effective both in radial as well as in axial directions. In a simpler
approach, however, we ignore the $z-$dependence and consider the metric
functions depending only on the radial coordinate. This amounts to
reflection symmetry in the $z-$direction by choosing a slice of constant $z-$%
surface.

Although we have started by establishing a generic framework, the sources of
our spacetimes considered here will be a line source for the Levi-Civita
(LC)\ metric with a cosmological constant $\Lambda $ \cite{LC}, which with
the right selection of parameters can be reduced to a cosmic string (CS) 
\cite{Trendafilova} or a black string (BS) \cite{Lemos} metric. These are
chosen deliberately simple enough to expose the role of asymmetry in a
cylindrical spacetime. Choosing different values of the parameters on
different sides of the throat gives rise to asymmetry in the TSW. The next
step is to address the stability analysis of an ATSW by assuming a
generalized fluid equation of state (EoS) after the perturbation. Once a
metric is perturbed, a new energy-momentum arises to counterbalance the
nonzero curvature terms on the left-hand-side of the Einstein equations.
Instead of a barotropic EoS for the fluid, we shall assume a more general
case defined by $p=p\left( \sigma ,a\right) $ \cite{Varela} in which $p$ is
pressure, $\sigma $ is energy density, $a$ is the time-dependent radius of
the TSW and $p\left( \sigma ,a\right) $\ represents a differentiable
function of its arguments. Such an EoS is more appropriate for the TSW
perturbations since it involves more degrees of freedom to be accommodated
in comparison with the standard barotropic fluid given by the representation 
$p=p\left( \sigma \right) $. Following the usual formalism of Lanczos \cite%
{Lanczos} and Israel \cite{Israel} apt for the thin-shells, we obtain the
forms of $p$ and $\sigma $\ in terms of the metric functions and their
derivatives. These are usually tedious enough for an analytical treatment,
however, we can reduce the equation to a simple second order one involving a
potential. Stability of the system emerges as a result of the sign of the
second derivative of the potential in terms of the radius of the shell. Our
task reduces to plot numerically the regions of the second derivative of the
potential that admits positive sign which will be our stability regions.
Depending on the tuned parameters involved, the stability region can be
larger which is interpreted as a more stable system. In our previous study 
\cite{Forghani}, with spherical symmetric configurations of ATSWs we had
enough examples to conjecture that asymmetry and stability are inversely
related. Our intention is to pursue a similar analysis in the case of a
cylindrically symmetric throat. To mention a particular example at this
stage in the CS case when we have a deficit angle (less than $2\pi $) or
surplus angle (more than $2\pi $) around the defect, stability argument and
level of asymmetry do not show a parallelism.

Organization of the paper is as follows. In section II we introduce our
general formalism. Particular choices of spacetimes and their stability
analysis are discussed in section III. We complete the paper with our
concluding remarks in section IV.

\section{The Generic Formalism}

To establish a framework which the spacetimes with cylindrical geometry in
general relativity fit into, we begin by initiating the metrics of the two
sides of the wormhole in their most general cylindrically symmetric form as%
\begin{equation}
ds_{i}^{2}=-A_{i}\left( r_{i}\right) dt_{i}^{2}+B_{i}\left( r_{i}\right)
dr_{i}^{2}+C_{i}(r_{i})d\phi _{i}^{2}+D_{i}(r_{i})dz_{i}^{2};\text{ \ \ }%
i=1,2,
\end{equation}%
where the metric functions $A_{i}\left( r_{i}\right) $, $B_{i}\left(
r_{i}\right) $, $C_{i}\left( r_{i}\right) $ and $D_{i}\left( r_{i}\right) $
are all functions of $r_{i}$. To construct the TSW, the Visser's standard
cut and paste procedure comes to help \cite{Visser}. The instruction is to
cut a submanifold from each spacetime given by $\mathcal{M}_{i}=\left\{
x_{i}|r_{i}\geq a>r_{hi},i=1,2\right\} $, and to bring them together at $%
\partial \mathcal{M}_{i}=\{x_{i}|r_{i}=a,i=1,2\}$ which is their common
timelike hypersurface. Herein, $r_{hi}$ is the radius of any possible
horizon in the $i^{th}$ spacetime. By this, one creates a geodesically
complete Riemannian manifold which connects the two spacetimes at their
shared boundary $\partial \mathcal{M}$; the so-called throat of the
wormhole. The time-dependent equation defining $\partial \mathcal{M}$ can
implicitly be written as

\begin{equation}
\mathcal{F}_{i}\left( r_{i},\tau \right) =r_{i}-a\left( \tau \right) =0;%
\text{ \ \ }i=1,2,
\end{equation}%
where $\tau $\ is the proper time on the throat.

Now that the wormhole is constructed, imposing the Israel junction
conditions \cite{Israel} on the metric and the curvature of the throat will
be the next step. Firstly, these conditions give a unique metric on the TSW
given by%
\begin{equation}
ds_{shell}^{2}=-d\tau ^{2}+C(a\left( \tau \right) )d\phi ^{2}+D(a\left( \tau
\right) )dz^{2},
\end{equation}%
\qquad where now $C(a\left( \tau \right) )$ and $D(a\left( \tau \right) )$
necessarily satisfy%
\begin{equation}
\left\{ 
\begin{array}{c}
C(a\left( \tau \right) )=C_{i}(r_{i}) \\ 
D(a\left( \tau \right) )=D_{i}(r_{i})%
\end{array}%
\right.
\end{equation}%
on the throat, while $\dot{t}_{i}^{2}=\frac{1+B_{i}\left( r_{i}\right) \dot{a%
}^{2}}{A_{i}\left( r_{i}\right) }$ in which an overdot stands for the
derivative with respect to the proper time.

Secondly, passing through the TSW from one side to another, there is a jump
in the extrinsic curvature tensor which is an implication of presence of a
matter field on the throat. This second condition is mathematically
expressed as the Lanczos equations \cite{Lanczos} 
\begin{equation}
\left[ K_{b}^{a}\right] -\delta _{b}^{a}\left[ K\right] =-S_{b}^{a};\text{ \
\ }8\pi G=1,
\end{equation}%
where $K_{b}^{a}$ and $K$ are the mixed extrinsic curvature and its trace on
the throat, respectively. A square bracket implies a jump in the quantity it
embraces i.e. $\left[ \Upsilon \right] =\Upsilon _{2}-\Upsilon _{1}$. In
this context, $S_{b}^{a}$ is the energy-momentum tensor belonging to the
fluid localized on the throat, given by%
\begin{equation}
S_{b}^{a}=diag\left( -\sigma ,p_{\phi },p_{z}\right) ,
\end{equation}%
where $\sigma $\ is the energy density, while $p_{\phi }$\ and $p_{z}$\ are
the surface pressures of the fluid along $\phi $\ and\ $z$, respectively.
Later, it will be demonstrated that this fluid does not satisfy the proper
energy conditions, and therefore is considered exotic.

In attempt to establish Eq. (5) explicitly for our general metrics, we begin
by the definition of the components of the covariant extrinsic curvature
tensor (for each spacetime separately) given in general relativity by

\begin{equation}
K_{ab}=-n_{\mu }\left( \frac{\partial x^{\mu }}{\partial \xi ^{a}\partial
\xi ^{b}}+\Gamma _{\alpha \beta }^{\mu }\frac{\partial x^{\alpha }}{\partial
\xi ^{a}}\frac{\partial x^{\beta }}{\partial \xi ^{b}}\right) ,
\end{equation}%
wherein,

\begin{equation}
n_{\mu }=\left( g^{\alpha \beta }\frac{\partial \mathcal{F}}{\partial
x^{\alpha }}\frac{\partial \mathcal{F}}{\partial x^{\beta }}\right) ^{-1/2}%
\frac{\partial \mathcal{F}}{\partial x^{\mu }}
\end{equation}%
are the spacelike $4-$normal components, $\Gamma _{\alpha \beta }^{\mu }$
are the Christoffel symbols compatible with the metric of each spacetime, $%
x^{\mu }=\left\{ t,r,\theta ,\varphi \right\} $ are the coordinates of the
bulk spacetimes and finally, $\xi ^{b}=\left\{ \tau ,\theta ,\phi \right\} $
are the coordinates on the TSW.

Having considered all these, with some manipulations, the Lanczos equations
for energy density and the pressures along $\phi $\ and\ $z$\ add up to%
\begin{equation}
\sigma =-\frac{1}{2}\left( \ln \left( CD\right) \right) ^{\prime
}\sum_{i=1}^{2}\left( \sqrt{\frac{1+B_{i}\dot{a}^{2}}{B_{i}}}\right) ,
\end{equation}%
\begin{equation}
p_{\phi }=\frac{1}{2}\sum_{i=1}^{2}\left[ \frac{2B_{i}\ddot{a}+\left( \ln
\left( A_{i}B_{i}D\right) \right) ^{\prime }B_{i}\dot{a}^{2}+\left( \ln
\left( A_{i}D\right) \right) ^{\prime }}{\sqrt{B_{i}\left( 1+B_{i}\dot{a}%
^{2}\right) }}\right] ,
\end{equation}%
and%
\begin{equation}
p_{z}=\frac{1}{2}\sum_{i=1}^{2}\left[ \frac{2B_{i}\ddot{a}+\left( \ln \left(
A_{i}B_{i}C\right) \right) ^{\prime }B_{i}\dot{a}^{2}+\left( \ln \left(
A_{i}C\right) \right) ^{\prime }}{\sqrt{B_{i}\left( 1+B_{i}\dot{a}%
^{2}\right) }}\right] .
\end{equation}%
Herein, a prime and an overdot imply a total derivative with respect to the
radius $a$\ and the proper time $\tau $, respectively. 

As can be observed from Eq. (9), based on the presumption stating that all
the metric functions are positive functions, $\sigma $ is negative definite
and therefore the fluid on the throat, at least does not maintain the weak
energy condition and hence is regarded exotic.

Furthermore, Eq. (9) can be rearranged in the form of an equation%
\begin{equation}
\dot{a}^{2}+V(a)=0,
\end{equation}%
where now the potential%
\begin{equation}
V(a)=\frac{1}{2}\left( \frac{1}{B_{1}}+\frac{1}{B_{2}}\right) -\left[ \frac{%
\left( \ln \left( CD\right) \right) ^{\prime }}{4\sigma }\left( \frac{1}{%
B_{1}}-\frac{1}{B_{2}}\right) \right] ^{2}-\left[ \frac{\sigma }{\left( \ln
\left( CD\right) \right) ^{\prime }}\right] ^{2}
\end{equation}%
is a radius-dependent potential, subject to the linear stability analysis as
following. With the assumption that there is an equilibrium radius $%
a_{0}(>r_{hi})$, where the wormhole is stable at, this potential can be
Taylor-expanded about the equilibrium radius $a_{0}$ where $V\left(
a_{0}\right) $ and $V^{\prime }\left( a_{0}\right) $\ necessarily become
zero. Therefore, having a radial perturbation applied on the throat such
that it does not unsettle the cylindrical symmetry, the sign of the second
derivative of this potential with respect to the radius, $V^{\prime \prime
}(a_{0})$, will decide whether the throat is stable or not at the presumed
equilibrium radius $a_{0}$. On the way attaining this, one must bear in mind
that the three expressions in Eqs. (9-11) are not independent of each other
and are related by two generic variable EoS%
\begin{equation}
\left\{ 
\begin{array}{c}
p_{\phi }=p_{\phi }(\sigma ,a) \\ 
p_{z}=p_{z}(\sigma ,a)%
\end{array}%
\right. .
\end{equation}%
This approach is the main theme of the Garcia-Lobo-Visser (GLV) method of
linear stability analysis used in many articles \cite{Garcia}.

As a final necessity to this section, we will be in search for a proper
energy equation that holds on the throat. Pursuing this, one may start by
performing a covariant derivative on the energy-momentum tensor $S^{ab}$ 
\cite{Mazhari}\ in the fashion%
\begin{equation}
\nabla _{b}S^{ab}\overset{a=\tau }{=}\sigma ^{\prime }+\frac{1}{2}\left[
\left( \ln \left( CD\right) \right) ^{\prime }\sigma +\left( \ln C\right)
^{\prime }p_{\phi }+\left( \ln D\right) ^{\prime }p_{z}\right] ,
\end{equation}%
which by direct substitution from Eqs. (9-11) yields%
\begin{multline}
\sigma ^{\prime }+\frac{1}{2}\left[ \left( \ln \left( CD\right) \right)
^{\prime }\sigma +\left( \ln C\right) ^{\prime }p_{\phi }+\left( \ln
D\right) ^{\prime }p_{z}\right] = \\
\sum_{i=1}^{2}\left\{ -\frac{1}{2}\left[ \left( \ln \left( CD\right) \right)
^{\prime \prime }+\frac{1}{2}\left( \ln \left( CD\right) \right) ^{\prime 2}%
\right] \left( \sqrt{\frac{1+B_{i}\dot{a}^{2}}{B_{i}}}\right) +\frac{\Xi }{4%
\sqrt{B_{i}\left( 1+B_{i}\dot{a}^{2}\right) }}\right\} .
\end{multline}%
where 
\begin{multline*}
\Xi \equiv \left( \ln \left( CD\right) \right) ^{\prime }\left( \ln
B_{i}\right) ^{\prime }+B_{i}\dot{a}^{2}\left[ \left( \ln C\right) ^{\prime
}\left( \ln \left( A_{i}B_{i}D\right) \right) ^{\prime }+\left( \ln D\right)
^{\prime }\left( \ln \left( A_{i}B_{i}C\right) \right) ^{\prime }\right] + \\
\left( \ln C\right) ^{\prime }\left( \ln \left( A_{i}D\right) \right)
^{\prime }+\left( \ln D\right) ^{\prime }\left( \ln \left( A_{i}C\right)
\right) ^{\prime }
\end{multline*}%
As it will be shown in the following sections, this energy equation will
play an important role on the way obtaining $V^{\prime \prime }(a_{0})$, and
therefore is vital to have the stability analysis accomplished.

Given a certain metric in the next section, Eqs. (9-11) and their static
counterparts, together with Eqs.(13) and (16) will be exploited for two
ATSWs, with the first being made by two cosmic string (CS) universes of
different deficit angles, namely a CS-CS* ATSW, and the second created by
connecting two black string (BS) geometries with different mass densities,
called a BS-BS* ATSW. There, we will write the energy relation in Eq.(16)
between $\sigma $, $p_{\phi }$\ and $p_{z}$ and getting assisted from, we
follow the GLV method with a variable EoS \cite{Varela} to see whether the
asymmetry favors among the more or less stable states.

\section{Stability Analysis of specific ATSWs}

As the subject matter, first we will have an overview on a rather general
non-rotating metric in cylindrical coordinates which is known under diverse
names; while some authors intend to call it the Levi-Civita (LC) solutions
with a non-zero cosmological constant (LCC or LC$\Lambda $) \cite{da Silva,
Zofka}, some others call it the Linet-Tian (LT) metric \cite{Brito, Eiroa}.
We will refer to it as the latter. If $\Omega >0$\ is the conicity of the
spacetime, this metric for a negative cosmological constant has a general
form of \cite{Linet, Tian}%
\begin{equation}
ds^{2}=dr^{2}+Q\left( r\right) ^{\frac{2}{3}}\left[ -P(r)^{\frac{2\mu \left(
\lambda \right) }{3\kappa \left( \lambda \right) }}dt^{2}+\frac{1}{\Omega
^{2}}P(r)^{\frac{2\nu \left( \lambda \right) }{3\kappa \left( \lambda
\right) }}d\phi ^{2}+P(r)^{\frac{2\xi \left( \lambda \right) }{3\kappa
\left( \lambda \right) }}dz^{2}\right] ,
\end{equation}%
where the functions%
\begin{equation}
\left\{ 
\begin{array}{c}
Q\left( r\right) =\frac{\sinh \left( \sqrt{-3\Lambda }r\right) }{\sqrt{%
-3\Lambda }} \\ 
P(r)=\frac{2\tanh \left( \sqrt{-3\Lambda }r/2\right) }{\sqrt{-3\Lambda }}%
\end{array}%
\right. 
\end{equation}%
include the cosmological constant $\Lambda $, and%
\begin{equation}
\left\{ 
\begin{array}{c}
\kappa \left( \lambda \right) =4\lambda ^{2}-2\lambda +1, \\ 
\mu \left( \lambda \right) =-4\lambda ^{2}+8\lambda -1, \\ 
\nu \left( \lambda \right) =-4\lambda ^{2}-4\lambda +2, \\ 
\xi \left( \lambda \right) =8\lambda ^{2}-4\lambda -1,%
\end{array}%
\right. 
\end{equation}%
are functions of $\lambda $ - a parameter related to the linear mass density
of the source \cite{da Silva} - and satisfy the constraint $\mu \left(
\lambda \right) +\nu \left( \lambda \right) +\xi \left( \lambda \right) =0$.
Due to the available symmetries \cite{Brito}, the conicity characteristics
of the spacetime and the behavior of geodesics \cite{Zofka}, the permitted
domain of $\lambda $ is $\left[ 0,1/2\right] $ \cite{Eiroa}. In order to
have the metric for a positive cosmological constant, however, one
substitutes the hyperbolic functions for their normal trigonometric
counterparts, and $-\Lambda $ for $\Lambda $ \cite{Zofka}. The metrics for
either positive or negative cosmological constant, recovers the LC solution
when $\Lambda \rightarrow 0$ \cite{LC}.

Nevertheless, we intend to focus more on\ the LT metric with $\Lambda <0$,
because this metric has similar properties to anti-de Sitter (AdS) spacetime
at $r\rightarrow \infty $\ when $\lambda $ is set to zero. This, however,
does not mean that for $\lambda =0$ we have exactly the AdS spacetime,
because AdS, when is written in cylindrical coordinates, is not static \cite%
{Bonner}. This makes this case more realistic compared to either the LC
solutions for $\Lambda =0$ or the LT solutions with $\Lambda >0$. It is also
worth mentioning that the solutions in Eq. (17) are not singular anywhere in
spacetime apart from the axis at $r=0$, for which even $r=0$\ is
non-singular when $\lambda =0$ or $\lambda =1/2$ \cite{Zofka}.\newline

When it comes to an ATSW made by two non-identical LT universes (an LT-LT*
ATSW), one must be sure that the conditions in Eq. (4) are satisfied. In
general, this explicitly means that the highly non-linear relations%
\begin{equation}
\frac{1}{\Omega _{1}^{2}}Q\left( r,\Lambda _{1}\right) ^{\frac{2}{3}%
}P(r,\Lambda _{1})^{\frac{2\nu \left( \lambda _{1}\right) }{3\kappa \left(
\lambda _{1}\right) }}=\frac{1}{\Omega _{2}^{2}}Q\left( r,\Lambda
_{2}\right) ^{\frac{2}{3}}P(r,\Lambda _{2})^{\frac{2\nu \left( \lambda
_{2}\right) }{3\kappa \left( \lambda _{2}\right) }}
\end{equation}%
\begin{equation}
Q\left( r,\Lambda _{1}\right) ^{\frac{2}{3}}P(r,\Lambda _{1})^{\frac{2\xi
\left( \lambda _{1}\right) }{3\kappa \left( \lambda _{1}\right) }}=Q\left(
r,\Lambda _{2}\right) ^{\frac{2}{3}}P(r,\Lambda _{2})^{\frac{2\xi \left(
\lambda _{2}\right) }{3\kappa \left( \lambda _{2}\right) }}
\end{equation}%
must hold simultaneously. With the level of complexity the two above
equations bring into the calculations, studying the stability of the ATSW
will practically be impossible. Instead, we will try to have a look on two
certain metrics which are generated from the rather general LT metric under
special conditions. Again, we emphasize that due to the conditions in Eq.
(4), not all the metrics that the LT metric includes can be subject to an
ATSW study. For example, the LC metric itself cannot be examined in the ATSW
context, because when the conditions $C(a\left( \tau \right) )=C_{i}(r_{i})$
and $D(a\left( \tau \right) )=D_{i}(r_{i})$ are satisfied for such a metric,
the TSW cannot be asymmetric anymore. Let us proceed with particular
examples.

\subsection{CS-CS* ATSW}

As it was discussed in the lines above, while $\Lambda \rightarrow 0$ in the
LT metric evokes the LC solutions, $\lambda \rightarrow 0$ gives rise to the
so-called non-uniform AdS metric \cite{Bonner}. Now, if one combines these
two limiting conditions, the metric 
\begin{equation}
ds^{2}=-dt^{2}+\Omega ^{2}d\rho ^{2}+\rho ^{2}d\phi ^{2}+dz^{2},
\end{equation}%
arises, which is reparametrized with%
\begin{equation}
\rho =\frac{r}{\Omega },
\end{equation}%
separately for the side universes. This metric which is a vacuum solution to
the Einstein equation is identified by many names; "\textit{the metric of a
straight spinning string in cylindrical coordinates with parameter }$a=0$"
in \cite{Perlick}, the cosmic string (CS) metric \cite{Trendafilova} or the
Gott's solution \cite{Gott}. Comparing Eqs. (1) and (22) and recalling
Eq.(4), we see that on the throat it appoints 
\begin{equation}
\left\{ 
\begin{array}{c}
A_{i}=1 \\ 
B_{i}=\Omega _{i}^{2} \\ 
C=a^{2} \\ 
D=1%
\end{array}%
\right. .
\end{equation}%
Besides, we would like to redefine $\Omega _{i}$ such that they are related
to each other by an asymmetry factor $\epsilon $\ so that we can investigate
our results based on the degree of asymmetry i.e. the value of $\epsilon $.
Hence, we would like to set down%
\begin{equation}
\left\{ 
\begin{array}{c}
\Omega _{1}=\Omega \\ 
\Omega _{2}=\left( 1+\epsilon \right) \Omega%
\end{array}%
\right. ,
\end{equation}%
where the legitimate domain of $\epsilon $ is $\left( -1,\infty \right) $.
It is of interest to note that while for $\Omega >1$ the spacetime has a
conical geometry at $t=const.$ and $z=const.$ with angular defect $\delta
=2\pi \left( \frac{\Omega -1}{\Omega }\right) $, for $\Omega <1$\ there
exists a surplus angle.

On the energy conservation, Eq.(16) takes the following simple form%
\begin{equation}
\sigma ^{\prime }+\frac{1}{a}\left( \sigma +p_{\phi }\right) =0,
\end{equation}%
in which there is no wake of $p_{z}$\ for the metric in every plane with $%
t=const.$ and $z=const.$ has the same geometry. Hence, the energy density
and the angular pressure given by Eqs. (9) and (10), and their static
counterparts reduce to%
\begin{equation}
\sigma =-\frac{1}{a}\left[ \frac{1}{\Omega }\sqrt{1+\Omega ^{2}\dot{a}^{2}}+%
\frac{1}{\left( 1+\epsilon \right) \Omega }\sqrt{1+\left( 1+\epsilon \right)
^{2}\Omega ^{2}\dot{a}^{2}}\right] ,
\end{equation}%
\begin{equation}
p_{\phi }=\ddot{a}\Omega \left[ \frac{1}{\sqrt{1+\Omega ^{2}\dot{a}^{2}}}+%
\frac{\left( 1+\epsilon \right) }{\sqrt{1+\left( 1+\epsilon \right)
^{2}\Omega ^{2}\dot{a}^{2}}}\right] ,
\end{equation}%
\begin{equation}
\sigma _{0}=-\frac{1}{a_{0}\Omega }\left( \frac{2+\epsilon }{1+\epsilon }%
\right) ,
\end{equation}%
and%
\begin{equation}
p_{\phi 0}=0,
\end{equation}%
on the throat, which are in complete agreement with the results in \cite%
{Eiroa2}. Accordingly, for a CS-CS* ATSW, the potential expression in Eq.
(13) will explicitly be%
\begin{equation}
V\left( a\right) =-\left( \frac{\sigma a}{2}\right) ^{2}-\left[ \frac{%
\epsilon \left( 2+\epsilon \right) }{2\sigma \left( 1+\epsilon \right)
^{2}\Omega ^{2}a}\right] ^{2}+\frac{1+\left( 1+\epsilon \right) ^{2}}{%
2\left( 1+\epsilon \right) ^{2}\Omega ^{2}}.
\end{equation}%
To compute $V^{\prime \prime }(a)$, we need the first and second derivatives
of the energy density with respect to the radius. Thus, wherever is needed,
for the first derivative of the energy density $\sigma ^{\prime }$ we
substitute from Eq.(26) and for the second derivative we will apply 
\begin{equation}
\sigma ^{\prime \prime }=\frac{1}{a^{2}}\left( \sigma +p_{\phi }\right)
\left( \omega _{2}+1\right) -\frac{\omega _{1}}{a},
\end{equation}%
in which we have used $\omega _{1}=\frac{\partial p_{\phi }\left( \sigma
,a\right) }{\partial a}$ and $\omega _{2}=\frac{\partial p_{\phi }\left(
\sigma ,a\right) }{\partial \sigma }$ in%
\begin{equation}
p_{\phi }^{\prime }=\frac{\partial p_{\phi }\left( \sigma ,a\right) }{%
\partial a}+\frac{\partial p_{\phi }\left( \sigma ,a\right) }{\partial
\sigma }\sigma ^{\prime }.
\end{equation}%
Eventually, by replacing the radius $a$ with a scaled radius $x=\Omega a$,
and $x_{0}=\Omega a_{0}$\ we manage to calculate the second derivative of
the potential at the equilibrium radius as%
\begin{equation}
V^{\prime \prime }(x_{0})=-2\left[ \frac{\omega _{2}}{\left( 1+\epsilon
\right) x_{0}^{2}}+\frac{\omega _{1}}{\left( 2+\epsilon \right) \Omega }%
\right] .
\end{equation}%
Solving $V^{\prime \prime }(x_{0})=0$ for $\omega _{2}$ leads to%
\begin{equation}
\omega _{2}=-\frac{\left( 1+\epsilon \right) x_{0}^{2}\omega _{1}}{\left(
2+\epsilon \right) \Omega },
\end{equation}%
which is the target equation for the final analysis. It is straightforward
to observe, that the angular pressure $p_{\phi }$ is either a function of
both the radius $a$ and the energy density $\sigma $\ or of none. Therefore,
a barotropic EoS cannot support a CS-CS* ATSW or a CS-CS TSW, at least, in
cylindrical coordinates. Additionally, $\omega _{2}$, which corresponds to
the speed of sound (say $c$) through the matter on the throat, is positive,
and then physically meaningful merely for the negative values of $\omega _{1}
$. Note that, since in this piece of work $c=1$, $\omega _{2}$ makes
physical sense in $\left[ 0,1\right) $. Nonetheless, entering a scale factor
through $\omega _{1}$ is always an option, and thus, only the general
behavior of $\omega _{2}$\ is of physical importance.

In Fig. 1, $\omega _{2}$ is plotted against $\frac{x_{0}^{2}\omega _{1}}{%
\Omega }$ for different values of $\epsilon $ and the stable regions are
marked. As a non-trivial result, the stability constantly increases with $%
\epsilon $ going from $-1$ to $\infty $. This not only shows that the
symmetric case ($\epsilon =0$) is not the most neither the less stable state
of the CS-CS* ATSW, but also indicates a dissimilar behavior approaching
symmetry; for negative/positive values of $\epsilon ,$\ stability
increases/decreases for more symmetrical states. This dissimilitude can be
associated with the spacetimes being more "conic" or more "surplus",
although, yet it may demand more scrutiny. 
\begin{figure}[tbp]
\includegraphics[width=80mm,scale=0.7]{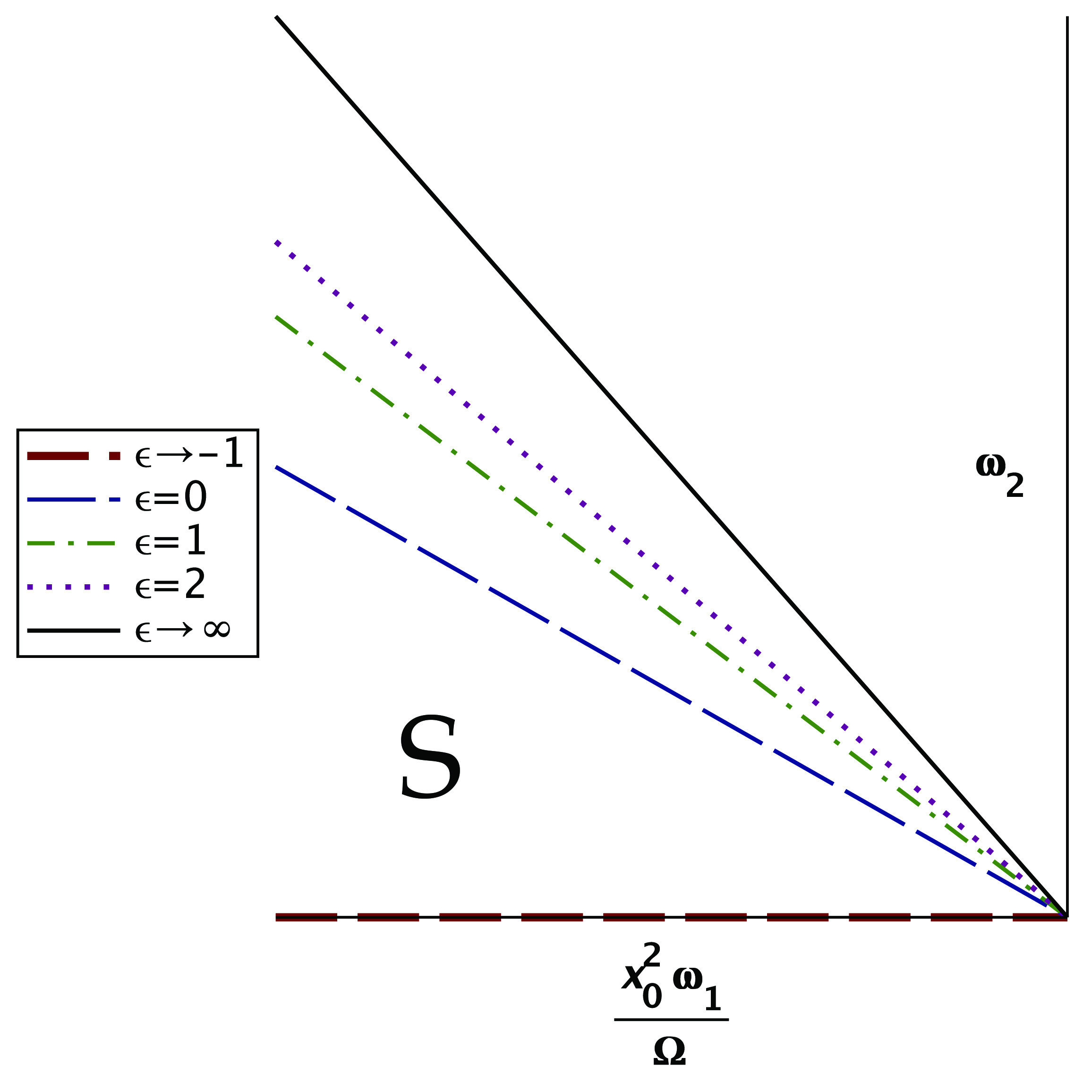}
\caption{{}Stability regions of the CS-CS* ATSW visualized by graphing $%
\protect\omega _{2}$ against $\frac{x_{0}^{2}\protect\omega _{1}}{\Omega }$
for various values of $\protect\epsilon \in \left( -1,\infty \right) .$ The
symmetric TSW corresponds to $\protect\epsilon =0.$ The stability increases
when $\protect\epsilon $ grows.}
\end{figure}

\subsection{BS-BS* ATSW}

Another interesting geometry can be constructed by setting $\lambda =\frac{1%
}{2}$ in Eq. (17) \cite{Zofka}. In a set of new coordinates $\left( T,\rho
,\varphi ,\zeta \right) $, the new metric, known under the name "\textit{The
uncharged, static black string metric}" \cite{Lemos}, is given by%
\begin{equation}
ds^{2}=-\Psi \left( \rho \right) dT^{2}+\frac{d\rho ^{2}}{\Psi \left( \rho
\right) }+\rho ^{2}d\varphi ^{2}+\alpha ^{2}\rho ^{2}d\zeta ^{2},
\end{equation}%
individually for the two spacetimes on the sides of the ATSW (for simplicity
the indices are removed), with 
\begin{equation}
\left\{ 
\begin{array}{c}
T=\frac{2\Omega }{-\Lambda }t \\ 
\rho =\frac{1}{\Omega }\cosh ^{\frac{2}{3}}\left( \frac{\sqrt{-3\Lambda }r}{2%
}\right) \\ 
\varphi =\phi \\ 
\zeta =\sqrt{\frac{3}{-\Lambda }}\Omega z%
\end{array}%
\right. ,
\end{equation}%
and the metric function%
\begin{equation}
\Psi \left( \rho \right) =\alpha ^{2}\rho ^{2}-\frac{4M}{\alpha \rho }.
\end{equation}%
Herein, the parameter $\alpha ^{2}=\frac{-\Lambda }{3}$, is positive
definite and $M=\frac{\alpha ^{3}}{4\Omega ^{3}}$ is associated with the
linear mass density of the black string calculated at radial infinity \cite%
{Brown}. This solution is non-singular and has a horizon at $\rho _{h}=\frac{%
\sqrt[3]{4M}}{\alpha }=\frac{1}{\Omega }$.

To investigate the stability of a BS-BS* ATSW, and for compatibility with
conditions in Eq. (4), we require that the two spacetimes possess the same
cosmological constant (and so the same $\alpha $) but different mass
densities. Hereinafter, the method and steps will be analogous to the
previous part's.

\begin{figure}[tbp]
\includegraphics[width=80mm, scale=0.7]{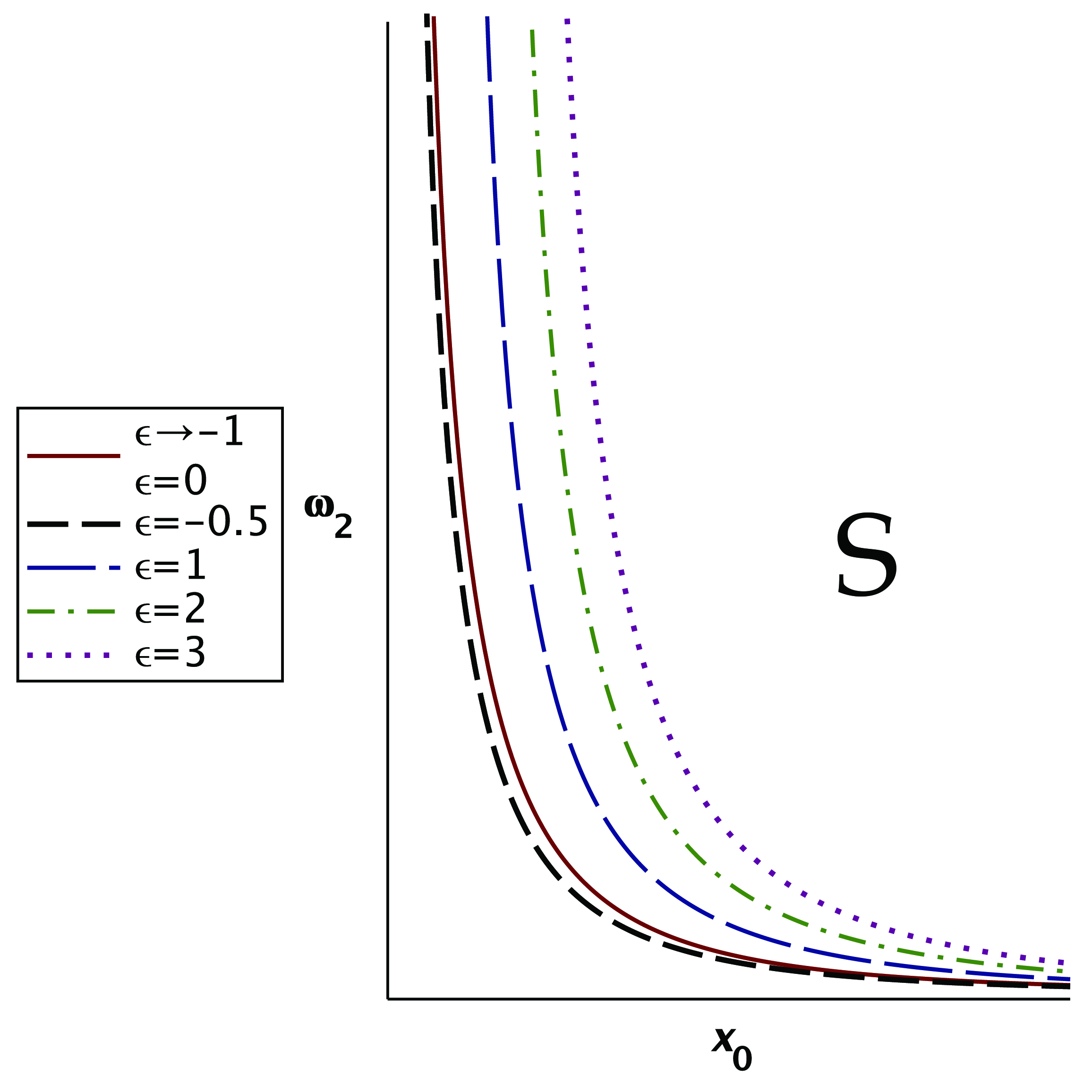}
\caption{{}Stability region of the BS-BS* ATSW\ against $x_{0}$ and $\protect%
\omega _{2}$ with a barotropic EoS, i.e., $\protect\omega _{1}=\protect%
\omega _{1a}=0$. Different curves correspond to different asymmetry factors $%
\protect\epsilon $. The symmetric TSW arises from $\protect\epsilon =0$
which is neither the maximum nor the minimum stable state.}
\end{figure}

Comparing Eqs. (1) and (36), we observe that on the throat%
\begin{equation}
\left\{ 
\begin{array}{c}
A_{i}=\Psi _{i}\left( a,M_{i}\right)  \\ 
B_{i}=\Psi _{i}^{-1}\left( a,M_{i}\right)  \\ 
C=a^{2} \\ 
D=\alpha ^{2}a^{2}%
\end{array}%
\right. ,
\end{equation}%
where we make the mass densities related to each other by an asymmetry
factor $\epsilon \in \left( -1,\infty \right) $ as follows%
\begin{equation}
\left\{ 
\begin{array}{c}
M_{1}=M \\ 
M_{2}=\left( 1+\epsilon \right) M%
\end{array}%
\right. .
\end{equation}%
Calculating the energy conservation given in Eq. (16), one obtains%
\begin{equation}
\sigma ^{\prime }+\frac{2}{a}\left( \sigma +p\right) =0.
\end{equation}%
in which we have regarded the fact that from Eqs. (10) and (11) the angular
and axial pressures on the throat are equal, i.e. $p_{\varphi }=p_{\zeta }=p$%
. By applying Eqs. (9) and (10) and the equalities in Eqs. (39) and (40),
the energy density and angular pressure and their static forms are%
\begin{equation}
\sigma =-\frac{2}{a}\left[ \sqrt{\alpha ^{2}a^{2}-\frac{4M}{\alpha a}+\dot{a}%
^{2}}+\sqrt{\alpha ^{2}a^{2}-\frac{4\left( 1+\epsilon \right) M}{\alpha a}+%
\dot{a}^{2}}\right] ,
\end{equation}%
\begin{equation}
p=\frac{2\left( \alpha ^{2}a^{2}-\frac{M}{\alpha a}\right) +a\ddot{a}+\dot{a}%
^{2}}{a\sqrt{\alpha ^{2}a^{2}-\frac{4M}{\alpha a}+\dot{a}^{2}}}+\frac{%
2\left( \alpha ^{2}a^{2}-\frac{\left( 1+\epsilon \right) M}{\alpha a}\right)
+a\ddot{a}+\dot{a}^{2}}{a\sqrt{\alpha ^{2}a^{2}-\frac{4\left( 1+\epsilon
\right) M}{\alpha a}+\dot{a}^{2}}},
\end{equation}%
\begin{equation}
\sigma _{0}=-\frac{1}{a_{0}}\left[ \sqrt{\alpha ^{2}a_{0}^{2}-\frac{4M}{%
\alpha a_{0}}}+\sqrt{\alpha ^{2}a_{0}^{2}-\frac{4\left( 1+\epsilon \right) M%
}{\alpha a_{0}}}\right] ,
\end{equation}%
and%
\begin{equation}
p_{0}=\frac{2\left( \alpha ^{2}a_{0}^{2}-\frac{M}{\alpha a_{0}}\right) }{a%
\sqrt{\alpha ^{2}a_{0}^{2}-\frac{4M}{\alpha a_{0}}}}+\frac{2\left( \alpha
^{2}a_{0}^{2}-\frac{\left( 1+\epsilon \right) M}{\alpha a_{0}}\right) }{a%
\sqrt{\alpha ^{2}a_{0}^{2}-\frac{4\left( 1+\epsilon \right) M}{\alpha a_{0}}}%
},
\end{equation}%
which in turn lead to the radial potential%
\begin{equation}
V\left( a\right) =\alpha ^{2}a^{2}-\left( \frac{\sigma a}{4}\right) ^{2}-%
\frac{2M\left( 2+\epsilon \right) }{\alpha a}-\left[ \frac{4\epsilon M}{%
\alpha \sigma a^{2}}\right] ^{2}.
\end{equation}%
\begin{figure}[tbp]
\includegraphics[width=80mm, scale=0.7]{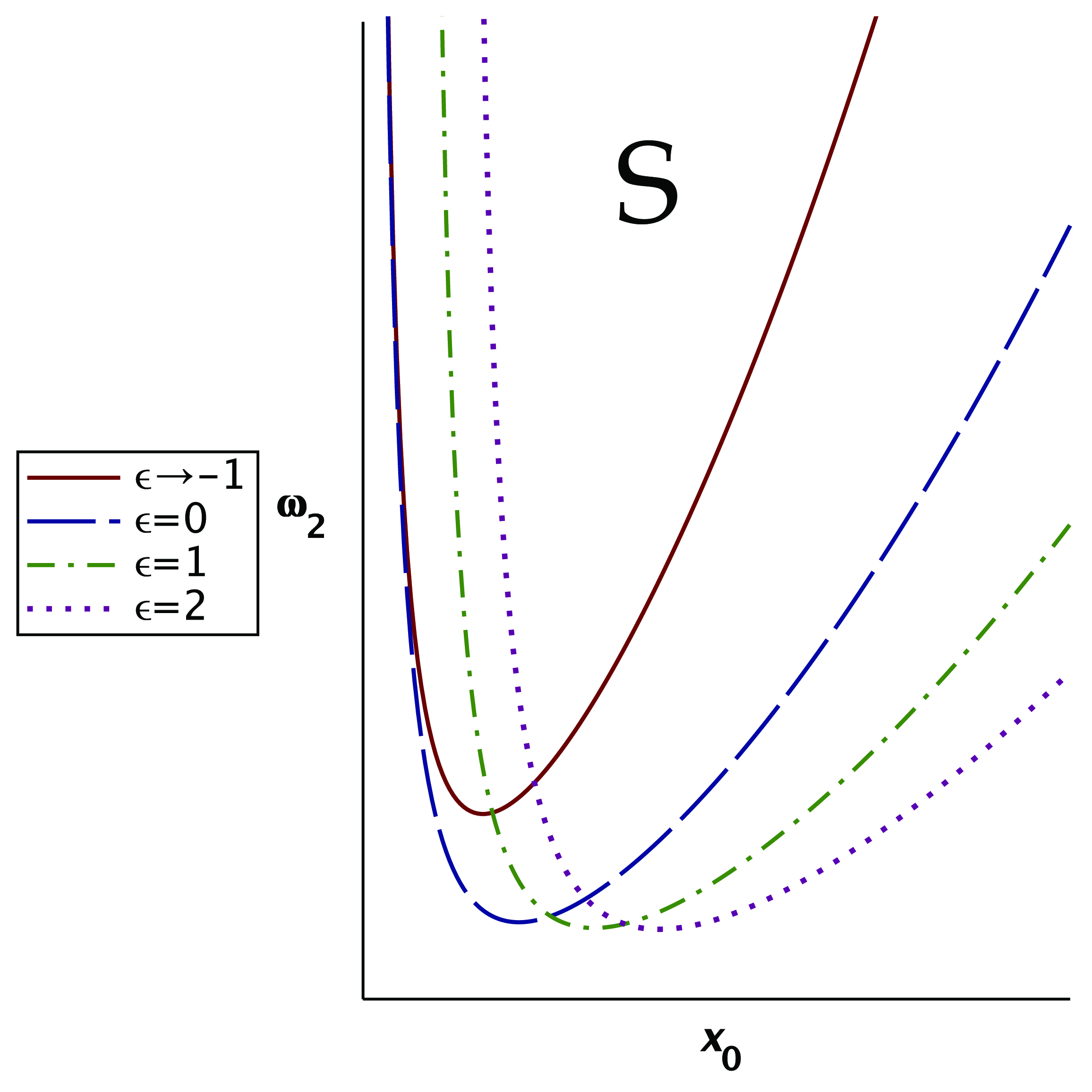}
\caption{{}Stability region of the BS-BS* ATSW supported by a variable EoS
with $\protect\omega _{1}=\protect\omega _{1b}=\protect\alpha ^{2}$. The
curves are drawn for various values of $\protect\epsilon $ including the
symmetric TSW with $\protect\epsilon =0.$ }
\end{figure}
\begin{figure}[tbp]
\includegraphics[width=80mm, scale=0.7]{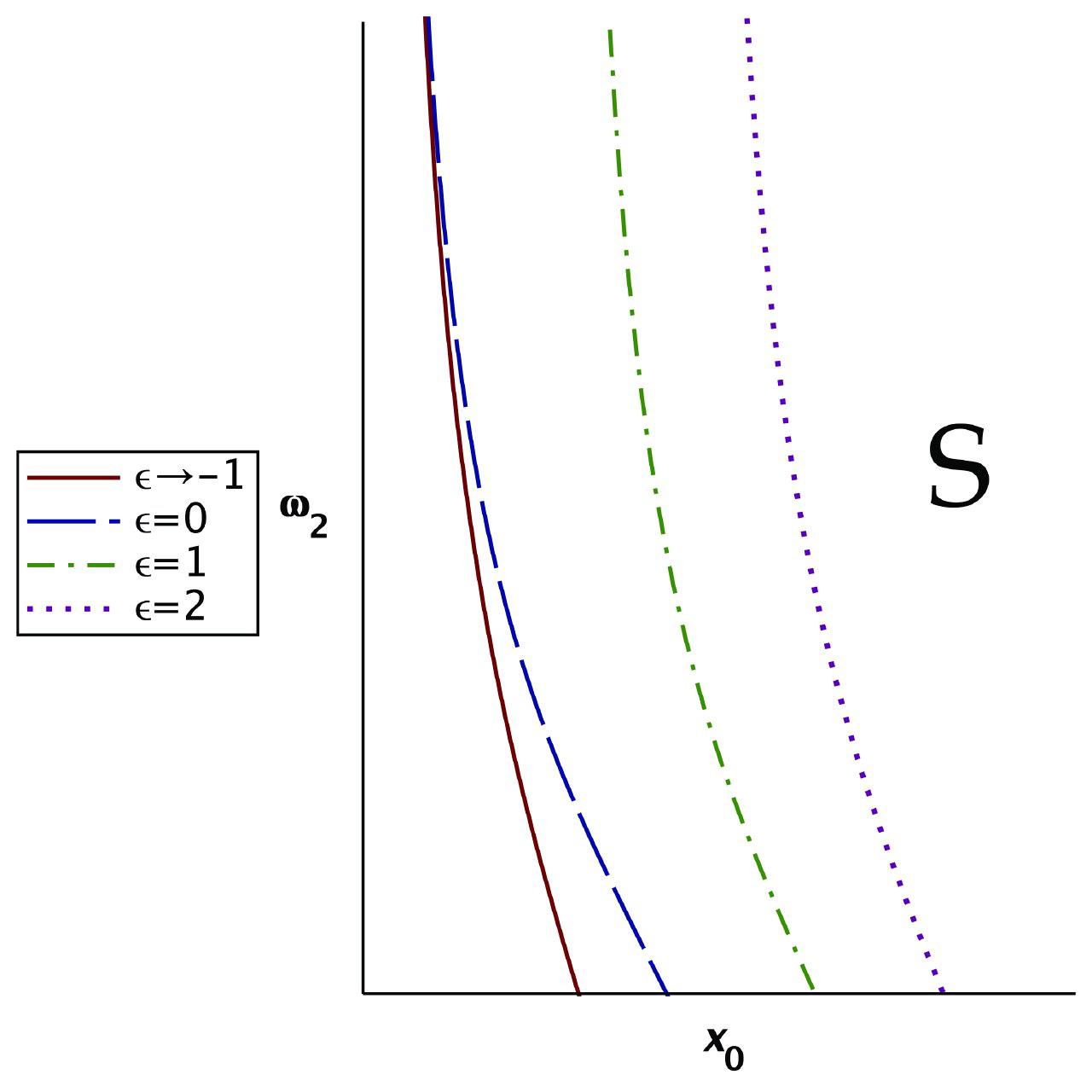}
\caption{{}Stability region of the BS-BS* ATSW obeying a variable EoS with $%
\protect\omega _{1}=\protect\omega _{1c}=-\protect\alpha ^{2}$. The graphs
are plotted for different values of $\protect\epsilon $, including $\protect%
\epsilon =0$ where a normal TSW emerges.}
\end{figure}

Having this potential, we can find $V^{\prime \prime }(x_{0})$, the second
derivative of the potential at the rescaled equilibrium radius $x_{0}=\alpha
a_{0}$, and from there, solving for $V^{\prime \prime }(x_{0})=0$ we manage
to extract an expression for $\omega _{2}$ in terms of $x_{0}$, $\epsilon $, 
$\alpha $, $M$\ and $\omega _{1}$. Here $\omega _{1}$ and $\omega _{2}$\
have the same definitions as the previous section in Eq. (32). To avoid
excessive sophistication, the explicit expressions for $V^{\prime \prime
}(x_{0})$ and $\omega _{2}$ are not brought here, although, the results are
graphed for $\omega _{2}$ against $x_{0}$ in Figs. 2, 3 and 4 for different
values of $\epsilon $. Without loss of generality we have set $M=1$\ in all
the figures, and regardless of a scale factor, only the general trend of the
graphs are intended to achieve. All the values on the $x_{0}-$axis are
beyond the horizon $x_{h}$. It is important to note that this horizon is
determined by the spacetime with the greater mass density; as far as $%
\epsilon $ lies in the domain $\left( -1,0\right] $, the horizon is at $%
x_{h}=\sqrt[3]{4}$, but once it exceeds zero, $x_{h}=\sqrt[3]{4\left(
1+\epsilon \right) }$.

For $\omega _{1a}=0$ the well-known barotropic EoS is summoned ($p=p(\sigma
) $). In this case, $\omega _{2}$ which is now also independent of $\alpha $%
,\ is plotted in Fig. 2 for $\epsilon \rightarrow -1$ and $\epsilon
=-0.5,1,2,3$. Here the catch is that $\epsilon \rightarrow -1$ and $\epsilon
=0$ lead to the same exact graph. The evolution of the graph by $\epsilon $
is such that starting from $\epsilon =0$ to the negative values stability
increases for a while (approximately up to $\epsilon =-0.5$), and then
decreases again so at $\epsilon \rightarrow -1$ it approaches the same
stability at $\epsilon =0$. for $\epsilon >0$, the mechanical stability
constantly reduces. Therefore, it seems that a BS-BS* ATSW with a barotropic
EoS \textit{is not} at its best stability when is symmetric, i.e. when $%
\epsilon =0$.

Figs. 3 and 4 are dedicated to two optionally selected constant variable
EoS, where%
\begin{equation}
\left\{ 
\begin{array}{c}
\omega _{1b}=\alpha ^{2} \\ 
\omega _{1c}=-\alpha ^{2}%
\end{array}%
\right. .
\end{equation}%
The motivation of bringing these examples with variable EoS is only to show
how greatly this type of EoS can alter the mechanical stability of an ATSW
and so is worth further study. The stable regions are marked with an $%
\mathsf{S}$.

\section{Conclusion}

Within the generic formalism, we have introduced\ classes of ATSWs in
cylindrical coordinates which are essentially less symmetric than their
spherical counterparts. As a result, the equality of the spherical angular
pressures, $p_{\theta }=p_{\varphi },$ split into different components $%
p_{\varphi }$ and $p_{z}$ in the cylindrical coordinates. The occurrence of
non-equal pressures naturally makes the problem more difficult, which is
simplified by relying only on the radial type of perturbations in the
stability analysis. The founded generic formalism in section II was applied
to the ATSW constructed by the LT bulk spacetime with three distinguishable
parameters $\Omega $, $\Lambda $ and $\lambda $. To find a way out of the
high level of complexity caused by the nature of the LT metric and the
conditions the study is subject to, we narrowed down the case studies to two
special metrics made of the LT metric under certain selections of its
parameters; the CS metric and the BS metric. Despite the apparent
differences, both CS-CS* ATSW and BS-BS* ATSW, were established by imposing
an asymmetry in the values of their side universes' conicity $\Omega $;
while this point is explicit for the former case, it is somehow less obvious
in the latter case where the asymmetry seemingly stems from a difference in
the side universes' values of the mass density $M$. However, these mass
densities are entangled with the conicities and cosmological constants of
the two spacetimes through $M_{i}=\left( -\Lambda _{i}/3\right)
^{3/2}/\left( 4\Omega _{i}\right) $, and since we required the same
cosmological constants ($\Lambda _{1}=\Lambda _{2}$) for the two spacetimes
due to conditions in Eq. (4), the difference in the mass densities is
actually coming from a difference in the conicities. Anyhow, the real
importance lies within the way the asymmetry is imposed; the value of the
asymmetric factor $\epsilon $.

Both from our physical intuitions and the experience we had gained from our
previous study on the spherical ATSWs \cite{Forghani}, we expected the
maximum stability of cylindrical ATSWs to coincide with their symmetries, as
well. Allegedly, this is not the case. Although there exists a pattern, it
was observed that for both ATSWs studied here, "the more symmetric" does not
necessarily admit "the more (or the less) stability". More detailed, in Fig.
1 we see that in CS-CS* ATSW, increasing $\epsilon $ from $-1$ to $\infty $
increases the stability. We recall that the symmetric TSW corresponds to $%
\epsilon =0$. So, deviating from symmetry in one way increases the
mechanical symmetry while in the other way decreases it. Also, for the case
of a BS-BS* ATSW, although the behavior of the stability region differs from
the former case, we see from Figs. 2-4 that the symmetric TSW is not the
most stable state.

\end{document}